\documentclass{article}
\newcommand{\bpi}{\mbox{\boldmath $\pi$}}

\begin{document}\begin{center}
{\large\bf Generalizations of MICZ-Kepler system}\\[4mm]

{\large Armen Nersessian}
\end{center}
{\sl
Yerevan State University, 1 A.Manoogian St., Yerevan, 375025 Armenia\\
 Artsakh State University, ,
  Stepanakert \&
 Yerevan Physics Institute,
 Yerevan,
 Armenia
}
\begin{abstract}
We discuss the generalizations of the  MICZ-Kepler
system (the system describing the motion of the charged particle in the field of Dirac dyon),
to the curved spaces, arbitrary potentials and to the multi-dyon
background.
\end{abstract}

 The integrable system describing the motion of the charged particle in the field of Dirac dyon (magnetic monopole carrying the electric charge)
     has been suggested  independently by Zwanziger \cite{Z}  and
      by McIntosh and Cisneros \cite{mic} and presently is known as MICZ-Kepler system. It is defined
by  the following
      Hamiltonian
      \begin{equation}
       {\cal H}_{\rm MIC}=\frac{\bpi^2}{2\mu} +\frac{{s}^2}{2\mu
      r^2}+\frac{\alpha}{r},\quad {\rm where }\quad
      [\pi_i,\pi_j]=-s\frac{\varepsilon_{ijk}x_k}{r^3},\quad
      [\pi_i,x_j]=-\imath\delta_{ij}.\label{1} \end{equation}
   The distinguished peculiarity  of the MICZ-Kepler system is the close similarity with Coulomb
      problem, which lie in the existence
   of a
      hidden symmetry given by the angular momentum operator ${\bf J}$
and by the analog of Runge-Lenz vector, which are defined by the
expressions
      \begin{equation}
{\bf J}={\bf r}\times \bpi +s\frac{{\bf r}}{r},\qquad
      {\bf
      A}=\frac{1}{2\mu}\left[\bpi\times{\bf J}-{\bf
      J}\times \bpi\right] +\alpha\frac{{\bf
      r}}{r} .\label{2} \end{equation}
This hidden symmetry exists due to the appearance, in  the
Hamiltonian, of the specific centrifugal term $s^2/2r^2$. The
origin of the additional centrifugal term $\frac{s^2}{2\mu r^2}$
can be  understood as follows \cite{terant}:
 In the presence of
monopole magnetic field  the angular momentum of the system gets
additional, spin-like term $s{\bf r}/r$. The magnetic
field of Dirac dyon  is ${\bf B}=g{\bf r}/{r^3}$. On the other
hand, there is linear relation between orbital and magnetic
momenta of single particle,
   ${\mathbf{{\mathcal{M}}}}=\frac{e}{2 \mu c
  }{\mathbf{J}}$.
Hence, the  interaction energy of this magnetic momentum with
magnetic field ${\bf B}$ is given by the expression:
  \begin{equation}
  U_B=-{\mathbf{{\mathcal{M}}}}{\mathbf{B}}=-\frac{e}{2 \mu
  c}{\mathbf{J}}\frac{g{\bf r}}{r^3}=\frac{s^2}{2 \mu r^2}, \label{UB}
  \end{equation}
 i.e. coincides with the centrifugal term in the MICZ-Kepler system \cite{terant}.
 Though, the additional part of the orbital
  momentum proportional to $s$ should rather be assigned to the
  electro-magnetic field than to the particle, this non-correct interpretation
  nevertheless leads to the correct expression.
Let us  mention, that MICZ-Kepler system also describes the
relative motion of two Dirac dyons with electric and magnetic
charges $(e_1, g_1)$ and $(e_2, g_2)$ with ${e_1g_2-e_2g_1}={ c}s
$. Similar to above mentioned case, the centrifugal term $s^2/2\mu
r^2$ could be interpreted as the interaction energy of the induced
dipole momentum with the dyon electric field plus interaction
energy of the induced magnetic moment with magnetic field (here
$\mu$ is the reduced mass).
 Notice, also, that this
term yields, when we treat to obtain the MICZ-Kepler system,
similar to the Coulomb system, from the four-dimensional
oscillator \cite{qmic,terant}. Let us mention, in this respect, that
 the Schroedinger
equation for the MICZ-Kepler system is equivalent to the
Schroedinger equation for the system of two well-separated BPS
monopoles/dyons (which possesses the Coulomb symmetry)
\cite{gibbons}.
The quantum mechanical spectra of the MICZ-Kepler and Coulomb systems looks also
quite similar to each other,
\begin{equation}
      {\cal E}=-\frac{\mu\alpha^2}{2n^2},\qquad
      n=n_r+l+1,\quad n_r=0,1,\ldots\end{equation}
The actual observable difference of the MICZ-Kepler system from the Coulomb one  lies in the
change of the range of validity of the total angular momentum
      from $l=0,1,\ldots$ to
      \begin{equation}
      l=|s|, |s|+1,  \ldots,\quad m=-l,\ldots, l
      \end{equation}
where  the monopole number $s$ takes (half-)integer values. This leads to
 the  $(2|s|+1)$-fold degeneracy of the ground state with respect to the azimuth quantum number $m$.

In some sense, quantum MICZ-Kepler system could be viewed as an exited hydrogen atom.
 However, upon perturbation by external fields, it gets qualitative differences.
For example, in the  linear potential field $e{\bf E}{\bf r}$ there is the linear Stark effect, similar to the hydrogen atom.
However, in contrast with the latter one, the linear Stark effect completely removes the degeneracy
 on the azimuth quantum number
\cite{mara,stark}, in contrast with the hydrogen atom.
Particularly, the ground state is given by the expression
 \begin{equation} {\cal
E}_{0}=-\frac{\mu\alpha^2}{2(|s|+1)^2}+ \frac{|e{\bf
E}|}{\mu\alpha}\; m\; {\rm sgn}\; s \;\left(|s|+\frac{3}{2}\right)
\end{equation} Also, considering the dipole transitions in the MICZ-Kepler
system under influence of planar electromagnetic wave, one can
observe the change of selection rules. Namely, in the conventional
spherically symmetric quantum mechanical  systems
the selection rules are given by
the expressions (see, e.g. \cite{ll})
 $
  m'=m, \;j'=j-1;\quad m'=m\pm 1\; ,j'=j\pm 1\;\quad m'=m\pm 1\;
,j'=j\mp 1\;.
  $
While
in the MICZ-Kepler system other transitions are also possible:
 \begin{equation}
m'=m, \;j'=j\;\quad m'=m\pm 1\; ,j'=j . \label{slm}\end{equation}
This is consequence of the violation of $P$-parity  by the
magnetic monopole, and is  the general feature  of the  quantum
mechanical systems with monopoles
 (see \cite{tomilchik,stark} and the references in \cite{comment}).
The specific effect of the choice  of the Coulomb potential is
hidden symmetry,
 essentially simplifying the analyses
      of the  system. For example, it makes possible  the separation of variables in few coordinate systems
      \cite{book}.
Few years ago the generalizations of the MICZ-Kepler system to the three-dimensional sphere \cite{kurochkin}
and two-sheet hyperboloid \cite{pogos} were suggested: the properties of these systems were also found to be quite similar
to the ones of the respective Coulomb systems. It was shown recently in our paper with L.Mardoyan and A.Yeranyan, that
 this similarity is not occasional, but is the general peculiarity
of the spherically symmetric systems with monopoles \cite{pla}.
Namely, if we put the Dirac monopole
 in the center of spherically symmetric system, and simultaneously change
the potential in accordance with the rule
 \begin{equation} U(r)\to
U(r)+\frac{{ s}^2}{2G(r)r^2}, \label{replacement}\end{equation}
  (where $ds^2=G(r)d{\bf r}d{\bf r}$,  $r=|{\bf r}|$), then the functional dependence of the energy spectrum on
  (radial and orbital) quantum numbers will not be changed.
The only change in the spectrum of such system is the
lifting of the low
        bound of the  range of validity of the orbital quantum number (and degeneracy of the ground state)
from $0$ to $|s|$.\\
 It is deductive, for illustration of these phenomena, to consider
  the classical trajectories of the generic spherically symmetric systems with Dirac monopole,
\begin{equation}
    {\cal H}=\frac{\bpi^2}{2G} +\frac{{s}^2}{2Gr^2}+U(r),\quad {\rm where }\quad
      \{\pi_i,\pi_j\}=-s\frac{\varepsilon_{ijk}x_k}{r^3},\quad
      \{\pi_i,x_j\}=\delta_{ij}\label{111} \end{equation}

The angular momentum of the system is given by the expression (\ref{2}),
from which  it follows
\begin{equation}
{\bf Jr}=sr,\quad {\bf J}^2={\bpi}^2 r^2-(\bpi{\bf r})^2=s^2.
\label{222}\end{equation}
 In order to find the trajectories of the
system, it is convenient
 to  direct
 the $x^3$ axis  along the vector ${\bf J}$,
i.e. to assume that $ {\bf J}=J_3\equiv J$. Upon this choice of
the coordinate system, we get, from the first equation
 \begin{equation}
\cos\theta\equiv\frac{ x^3}{ r}=\frac sJ.
\label{tan}\end{equation} From  the second expression in
(\ref{222}) we get
\begin{equation}
{\bpi}^2=p^2_r+\frac{{\bf J}^2-s^2}{r^2},\qquad p_r=\frac{\bpi{\bf r}}{r}
\label{pr}\end{equation}
Taking into account this expression, we can represent the Hamiltonian (\ref{111})
 in the following form
\begin{equation}
 {\cal H}=\frac{p^2_r}{2G} +\frac{{\bf J}^2}{2Gr^2}+U(r),\quad {\rm where}\quad p_r=G\frac{dr}{dt}
\end{equation}
 Also, we can introduce the angle
\begin{equation}
 \phi={\rm arctan} \frac{ x^1}{ x^2}\;,\quad
  \frac{d\phi}{dt}=\frac{2J}{G r^2},\label{teq}
\end{equation}
and observe, that the time evolution of $\phi$ and $r$, as well as the form of trajectory, $r=r(\phi)$
 are independent on monopole number $s$. The only impact of the magnetic monopole is the rotation
  of the orbital plane on the  finite angle $\cos\theta=s/J$ breaking the $P$-symmetry of the system,
   and the lifting of the lower bound of the orbital momentum
  (see Eq.(\ref{222})),
  \begin{equation}
  |{\bf J}|\geq |s|.
  \end{equation}
Hence, following the rule (\ref{replacement}), we can construct the ``MICZ-analog" of any
integrable  spherically-symmetric system, and present its explicit
classical and quantum mechanical solutions!

Nevertheless, the ``invariant meaning"  of the magic centrifugal term $s^2/2gr^2$ has remained unclear yet,
so that we can't still construct the ``MICZ-analogs" of the non-spherically symmetric, but, nevertheless,
integrable systems. While numerous  important mechanical systems have no spherical symmetry. Among them the most known are
two-center Kepler systems \cite{ll}. Let us remind, that problem of particle moving in the field of two Coulomb
    centers (or two-center Kepler problem) was solved in  the
    middle of XIX century by Jacobi. He established the
    integrability of two-center Kepler system and of its limiting case
    when one of the forced centers is placed at infinity which
    yields the homogeneous potential field,  in elliptic and
    parabolic coordinates respectively. The
    generalization of this picture to case where Coulomb centers
    are replaced by dyons has been proposed in our recent paper with S.Krivonos  and V. Ohanyan \cite{MICZ}.
    There  we proposed the generalization of the
    MICZ-Kepler replacement
    (\ref{replacement}) which can be used in case of $N$ Dirac monopoles,
 \begin{equation}
 U\left({\bf r}\right)\rightarrow \frac{1}{2G}\left( \frac{s_1}{r_1}+...+\frac{s_n}{r_n}
 \right)^2+U\left({\bf r} \right), \label{multMICZ}
 \end{equation}
where $s_i=eg_i$, with  $g_i$ be the magnetic charge of the $i$-th
monopole located at the point  with coordinates ${\bf a}_i$, and
$r_i=|{\bf r}-{\bf a}|$.

This replacement has the following  important features:
\begin{itemize}
\item The  system (without monopoles) admitting separation of
variables in elliptic/parabolic coordinates results in the
separable two-center
 MICZ-system (\ref{multMICZ}) with the Dirac monopoles placed at the foci
 of elliptic/parabolic coordinates.
\item
The system  admits the  ${\cal N}=4$ supersymmetric extension at
the following choice of  potential \begin{equation}
U_{0}=\frac{\kappa}{G}\left(\sum_{I}\frac{g_I}{|{\bf r}-{\bf
a}_I|} + {\bf B}_0{\cdot\;\bf r}\right)+\frac{\kappa^2}{2G}.
\label{sup} \end{equation} The respective supersymmetric system
has been constructed by Ivanov and Lechtenfeld in 2003 \cite{IL}.
\end{itemize}
Hence, in the case of the Euclidean  metric, $G=1$, this replacement
leads the well defined two-center MICZ-Kepler system. Also, in Euclidean space such a multi-center
MICZ-Kepler system admits the ${\cal N}= 4$ supersymmetric extension, if the electric and magnetic charges of
background dyons has the same ratio, i.e.  obey trivial
 Dirac-Schwinger-Zwanziger
quantization condition  \begin{equation}
g_i q_j - g_j q_i=0. 
\label{DSZ}\end{equation} In that case the potential of the
Euclidean multi-center MICZ-Kepler system belongs to the class
(\ref{sup}) (up to unessential constant).

However, this procedure fails in case of the multi-center Coulomb systems on
spheres and two-sheet hyperboloids, when the conformal flat metric and the potential
are defined, respectively, by the expressions
\begin{equation} ds^2=\frac{4r^2_0(d{\bf
r})^2}{(1+\epsilon r^2)^2}. \label{met} \end{equation} and
\begin{equation} U({\bf r})=e\phi_{q_1,\ldots,
q_N}=e\sum_{i=1}^N q_i\phi (|{\bf r}-{\bf a}_i|).
\label{sC}\end{equation} Here
\begin{equation}
 \phi (r)=
\frac{1}{2r_0}\frac{1-\epsilon { r}^2}{{ r}}
\end{equation}
is the Coulomb potential on the sphere ($\epsilon=1$) and two-sheet hyperboloid ($\epsilon=1$) \cite{sch}.

The reason is the following:
 due to the
existence of hidden symmetries given by Runge-Lenz vector the
(one-center) Coulomb system on  Euclidean space admits the separation of
variables in spherical, elliptic and parabolic coordinate.
 The Coulomb system on (pseudo)spheres also has the hidden symmetries given by the analog of Runge-Lenz vector.
Analogously  to Euclidean case, this hidden symmetry is connected
to the the separation  variables in few coordinate systems. These
coordinate systems turn to the spherical, elliptic and parabolic
ones at the planar limit.
 The connection  between   spherical and Cartesian  coordinates on the sphere
  is identical  with that on the Euclidean space, whereas the discrepancy is appeared for the elliptic
  and parabolic ones (see, e.g., \cite{bogush} and refs therein).
  Therefore, the statement of \cite{MICZ}  concerning
 the separability of variables for  ``MICZ-extended"  systems
 is no more valid for the Coulomb systems  on spheres and two-sheet hyperboloids.
  Moreover, the potential of the multi-center Coulomb  system
on (pseudo)sphere,
 does not belong to the class (\ref{sup}), admitting the ${\cal N}=4$ supersymmetric extension.

For the construction of the well-defined generalization multi-center MICZ Kepler system on sphere and other
rotationally  invariant curved spaces, we have to understand, where the centrifugal term
in (\ref{multMICZ}) came out from.

For this purpose let us remind, that Coulomb potentials on the
three-dimensional  Euclidean space, sphere and pseudosphere are
nothing else, but the rotationally invariant Green functions for
the Laplasians defined by the corresponding metrics, $\Delta
\phi_{C}=\delta({\bf r})$. While the vector potential of the Dirac
monopole is just one-form  dual to this Green function.
Explicitly, \begin{equation} \ast d{A}_D =-d\phi_C,\quad
\Rightarrow \quad\Delta\phi_C=\delta({\bf r}),\quad
\Delta\equiv\ast d\ast d+ d\ast d\ast \label{hodge}\end{equation}
Taking into account the duality between vectors and one- and
two-forms in the three-dimensional spaces, one can write down
these expressions in the following way
 \begin{eqnarray}
-\mathbf{\nabla}\phi_C=\mbox{rot} \mathbf{A}_{D},
\quad\Rightarrow\quad \Delta\phi_C =\delta({\bf r}).
\label{vmonop}
  \end{eqnarray}
Let us write down the explicit $so(3)$ invariant solutions of this
equations  on the  $so(3)$ invariant spaces. The  (zero-, one- and two-)form fields are
independent on the metric, hence, one
 has to choose
\begin{equation} d{\bf A}_D({\bf r})=\frac{({\bf r}\times d{\bf r})\wedge d{\bf
r}}{2r^3},\qquad d\phi(r)=\frac{d\phi(r)}{dr}\frac{{\bf r}d{\bf
r}}{r}. \end{equation} On the other hand the rotationally
invariant metrics can always be represented in the conformal-flat
form $ds^2=G(r)(d{\bf r})^2 $.
 Taking into account the conformal flatness of the
metrics, we get, from (\ref{hodge}) \begin{equation}
\frac{d\phi_C}{dr}=-\frac{1}{r^2\sqrt{G(r)}},\qquad
\Rightarrow\qquad \phi_C=-\int\frac{dr}{r^2\sqrt{G(r)}}.
\label{GC}\end{equation} Particularly, for the Euclidean space one
obtains $\phi_C=1/r$, and for the sphere and two sheet hyperboloid
(with the metrics (\ref{met}) the potential
 given by  (\ref{sC}).

Then we  define the multi-center MICZ-Kepler system on the $so(3)$
invariant space by the expression \begin{equation}
  \mathcal{H}=\frac{(\mathbf{p}-e \mathbf{A}_{g_1,\ldots,g_N})^2}{2G}+\frac{e^2
  \phi^2_{g_1,\ldots, g_N}}{2}+e\phi_{q_1,\ldots,q_N}, \label{iv1}
\end{equation} where $\mathbf{A}_{g_1,\ldots,g_N}=\sum_i g_i\mathbf{A}_D({\bf
r}-{\bf a}_i)$, $\phi_{q_1,\ldots,q_N}=\sum_i q_i\phi_C(r_i)$.

When the electric and magnetic charges of background dyons obey
the condition (\ref{DSZ}), or, equivalently, $q_i/g_i =e\kappa $,
this Hamiltonian can be represented in the form \begin{equation}
  \mathcal{H}=\frac{(\mathbf{p}-e \mathbf{A}_{g_1,\ldots,g_N})^2}{2G}+\frac{e^2(\phi_{g_1,\ldots, g_N}+\kappa/e)^2}{2}
  -\frac{\kappa^2}{2} \label{iv2}.
\end{equation} where ${\rm grad} (\phi_{g_1,\ldots, g_N}+\kappa/e)=-{\rm rot}
\mathbf{A}_{g_1,\ldots,g_N}$. This Hamiltonian admits the ${\cal
N}=4$ supersymmetric extension on the Euclidean space \cite{IL}
 and on the sphere  \cite{BK,BKO} (provided  the unessential constant $\kappa^2/2$ is omitted).

Actually, the bosonic part of the supersymmetric system on sphere constructed by Bellucci, Krivonos and Shcherbakov
 \cite{BK}  was the key ingredient, which
led us to postulate the Hamiltonian
of multi-center
MICZ-Kepler system on $so(3)$ invariant spaces (\ref{iv1}).
In our knowledge, the  supersymmetric extensions of the MICZ-Kepler system
on the generic conformal-flat space are unknown. However, it seems, that we
believe, that with the explicit component expressions for the
supercharges of the system on sphere \cite{BKO},
we will be able to construct it in the Hamiltonian framework.

 Notice, that in the case of one background dyon the MICZ-Kepler  Hamiltonian on the sphere and
 two-sheet hyperboloid coincides with the MICZ-Kepler
 Hamiltonian constructed within the rule (\ref{multMICZ}) up to unessential constant $\epsilon s^2/4r^2_0$.
 Hence, it is separable in the spherical coordinates, and in the modified
  elliptic and parabolic coordinates considered in Ref. \cite{bogush}.
  It seems, that two-center MICZ-Kepler system on
sphere and hyperboloid is also separable in these coordinates.\\

Let us conclude our consideration by the following speculation. The one-center MICZ-Kepler system  describes not only the
motion in the field of  fixed dyon, but also the relative motion of two dyons in the centre-of-mass system.
Similarly, by the use of the replacement (\ref{replacement}) we can equip the two-body system with the magnetic charges,
without changing the qualitative properties of the initial system. Similar phenomenon  was observed since its inception
in the Calogero model \cite{calogero} describing the $N$ one-dimensional identical particles
with pair-wise inverse-square  and harmonic interaction: the energy spectrum of Calogero system coincides with the one of
$N$ free harmonic oscillators, apart from the coupling dependent shift of the ground-state energy
(for the recent review see \cite{polychronakos}). This structure of
spectrum has led Calogero to suggest the possibility of a map between Calogero model and the system
 of decoupled oscillators, which has been found very recently \cite{GP} (see also \cite{GLP}).
 Taking in mind the mentioned similarity between two-body MICZ-Kepler system and Calogero system, as well
 as the structure of multi-center MICZ-Kepler system, one can suppose, that the following
system of $N$ interacting monopoles could be viewed as a three-dimensional generalization of the Calogero model,
$${\cal H}=\frac{1}{2}\sum_{I}\left({\bf p}_I-\sum_Js_{IJ}A_{D}({\bf r}_I-{\bf r}_J)\right)^2+
\frac{1}{2}\sum_{\begin{array}{c}I,J,K\\I\neq J, I\neq K\end{array}}\frac{s_{IJ}s_{IK}}{r_{IJ}r_{IK}}+
\frac{1}{2}\sum_{I}\omega^2{\bf r}^2_IU,
$$
where $ r_{IJ}=|{\bf r}_I -{\bf r}_J|$, $s_{IJ}=e_Ig_J-e_Jg_I$.
But we didn't checked  this claim up to now.\\

{\large Acknowledgements.} I am indebted to  Sergey Krivonos,
Levon Mardoyan, Vadim Ohanyan, Armen Yeranyan for collaboration in
problems  reflected in this paper. I am grateful to the Organizers
of  Colloquium  on {\sl Integrable models and Quantum symmetry}
 and personally Professor Cestmir Burdik
for kind hospitality in Prague. This work is  partially supported
upported by the grants
 NFSAT-CRDF  ARP1-3228-YE-04
 and  INTAS-05-7928

\end{document}